\documentclass[pra,superscriptaddress,twocolumn,a4paper]{revtex4}

\usepackage{bm}
\usepackage{graphicx}
\usepackage{amsmath,amssymb}
\usepackage{hyperref}
\usepackage[latin1]{inputenc}

\newcommand{\figwidth}{0.99\columnwidth}

\newcommand{\tr}{\operatorname{tr}}

\newcommand{\kB}{k_\text{B}}

\begin{document}

\title{Finite-temperature properties of one-dimensional hard-core bosons in a quasiperiodic optical lattice}
\author{Nicolas Nessi}
\author{Aníbal Iucci}

\affiliation{Departamento de F\'{\i}sica, Facultad de Ciencias
Exactas, Universidad Nacional de La Plata and IFLP-CONICET, CC 67,
 1900 La Plata, Argentina.}

\begin{abstract}
We investigate the properties of impenetrable bosons confined in a one-dimensional lattice at finite temperature in the presence of an additional incommensurate periodic potential. Relying on the exact Fermi-Bose mapping, we study the effects of temperature on the one-particle density matrix and related quantities such as the momentum distribution function and the natural orbitals. We found evidence of a finite temperature crossover related to the zero temperature superfluid-to-Bose-glass-transition that induces a delocalization of the lowest natural orbitals.
\end{abstract}


\maketitle

\section{Introduction}

Recent developments in the field of ultracold atomic gases have opened up the possibility to investigating various challenges from the physics of strongly correlated systems~\cite{bloch08_cold_atoms_optical_lattices_review}. The remarkable experimental control over most of the system parameters, including dimensionality and interactions, makes it possible to use these setups as ``quantum simulators" of fundamental Hamiltonian models, otherwise inaccessible in more traditional condensed-matter realizations. For instance, the achievement of strong interactions and confinement to low-dimensional geometries led to the observation of the Tonks-Girardeau gas, with the addition of an optical lattice along the one-dimensional axis~\cite{paredes04_tonks_gas} and in the continuum~\cite{kinoshita04_1D_tonks_gas}.

A specially active line of research is the study of disorder-related phenomena~\cite{sanchez-palencia10_disorder_under_control}. One of the possibilities to create random potentials in a controlled way is to use laser beams generating speckle patterns~\cite{lye05_speckle_disorder,clement05_speckle_transport,fort05_speckle_expansion,%
schulte05_anderson_localization,chen08_phase_coherence,white09_disordered_bosons_optical_lattice}. An alternative technique is to superimpose two optical lattices with incommensurate frequencies, thus generating a quasiperiodic potential~\cite{fallani07_bichromatic_shaking}. These realizations of ultracold atoms in disordered potentials enabled the observation of Anderson localization in the regime of negligible interactions~\cite{billy08_observation_anderson_localization,roati08_anderson_localization_BEC}. However, strictly speaking, Anderson localization is a single-particle effect, and hence the quest turned to the understanding of the interplay between disorder and interactions, a particularly intriguing aspect of the localization phenomenon on which recent experiments on bosonic systems~\cite{deissler10_bosons_disorder_interactions,pasienski10_bosons_disorder_interactions} were performed. From the theoretical point of view, interacting bosons in a disordered potential are expected to exhibit a phase transition from a superfluid phase with extended single-particle states to a Bose-glass phase in which single-particle wavefunctions are Anderson localized as the interaction or the disorder strength is augmented\cite{giamarchi88_anderson_localization_bose_glass_1D,fisher89_bose_glass}. However, in spite of the theoretical efforts, the ground-state properties of disordered interacting bosons are still under debate~\cite{roscilde08_one_dimensional_superlattices,roux08_quasiperiodic_phase_diagram,deng08_phase_diagram_bichromatic,%
gurarie09_phase_diagram_disordered_BH,pollet09_phase_diagram_disorder_BH,orso09_shaking_disorder}, and, for instance, the possibility of a reentrant superfluid induced by disorder~\cite{sanchez-palencia10_disorder_under_control} was experimentally excluded~\cite{pasienski10_bosons_disorder_interactions}. A finite temperature was proposed as a source of discrepancies~\cite{pasienski10_bosons_disorder_interactions}. The reason for this lack of agreement lies in the high complexity of the models analyzed such as the disordered Bose-Hubbard model, which can only be solved by means of sophisticated numerical techniques. In this paper we concentrate on the finite-temperature properties of interacting bosons confined on a one-dimensional quasiperiodic potential in the limit of infinite repulsion, for which an exact solution exists~\cite{cazalilla11_1D_bosons_review} based on a mapping of the Hamiltonian to spinless fermions through the Jordan-Wigner transformation. We focus on the effect of the temperature on the behavior of off-diagonal one-particle correlations and related quantities such as the momentum distribution and the natural orbitals and their occupations. We analyze the finite-temperature crossover in the vicinity of the transition from the superfluid to the Bose glass at a finite value of the strength of the secondary potential. We find that the lowest natural orbitals, which are localized on the Bose-glass phase at zero temperature, delocalize when the system enters into the quantum critical region.

\section{Model and Methods}

We consider hard-core bosons in a one-dimensional lattice subjected to a quasiperiodic potential, described by the following Hamiltonian:
\begin{equation}\label{eq:hamhcb}
H=-t\sum_{i=1}^{L}(b_{i+1}^{\dag}b_{i}+\mathrm{h.c.})+\Delta\sum_{i}\cos(2\pi i\sigma)n_{i},
\end{equation}
where the boson destruction and creation operators $b_{i}$ and $b_{i}^{\dag}$ satisfy the special commutation relations
\begin{align}\label{eq:hcb_conmut}
\{b_{i},b_{i}^{\dagger}\}  &  =1,\quad\quad b_{i}^{2}=(b_{i}^{\dagger})^{2}=0,\\
[b_{i},b_{j}^{\dagger}]  &  =[b_{i},b_{j}]=[b_{i}^{\dagger},b_{j}^{\dagger}]=0,\quad\quad i\neq j,
\end{align}
avoiding multiple occupation of a site. $\Delta$ determines the strength of the incommensurate potential, and $\sigma$ is an irrational number that characterizes the degree of incommensurability. Throughout this paper we treat the case of $\sigma=(\sqrt{5}-1)/2$, which is a common choice.

The hard-core bosons Hamiltonian is exactly solvable by means of the Jordan-Wigner transformation,
\begin{equation}
c_{m}=\exp\left[-i{\pi}\sum_{j=1}^{m-1}n_{j}\right]b_{m},
\end{equation}
that maps Eq. (\ref{eq:hamhcb}) onto an equivalent noninteracting fermionic Hamiltonian,
\begin{equation}\label{eq:fercuasids}
    H_F=-t\sum_{i}^L(c_{i+1}^{\dag}c_{i}+\mathrm{h.c.})+\Delta\sum_{i}\cos(2\pi i\sigma)n_{i},
\end{equation}
where the $c$ and $c^{\dag}$ satisfy the canonical anticommutations relations. The boundary conditions must be handled with care since the last term of the sum over sites maps to $b_{L}^{\dag}b_{1}=-c_{L}^{\dag}c_{1}e^{i{\pi}\hat{N}}{\neq}c_{L}^{\dag}c_{1}$, with $\hat{N}$ the operator representing the total number of bosons. Thus, an even (odd) number of bosons induces antiperiodic (periodic) boundary conditions over the fermion model, which makes difficult its study in the grand-canonical ensemble. However, open boundary conditions are preserved by the mapping, and we consider this case in the present work.

We are interested in the properties of the one-particle density matrix, defined as
\begin{equation}\label{eq:opdm}
\rho_{ij}=\langle b_{i}^{\dag}b_{j}\rangle.
\end{equation}
At zero temperature $\langle\ldots\rangle$ represents a $N$-particle ground state expectation value. For this case, Rigol and Muramatsu~\cite{rigol04_emergence_quasicondensates_HCB,rigol04_lattice_hcbosons,rigol05_groundstate_hcbosons} found a very convenient way of representing Eq. (\ref{eq:opdm}), well suited for a numerical computation. The one-particle density matrix can be written as $\rho_{ij}=G_{ij}+\delta_{ij}(1-2G_{ii})$, where $G_{ij}$ is the one-particle Green's function and, according to Refs.~\onlinecite{rigol04_emergence_quasicondensates_HCB,rigol04_lattice_hcbosons,rigol05_groundstate_hcbosons}, can be represented in the form
\begin{equation}\label{eq:zero_t}
G_{ij}=\langle b_i b_j^{\dag}\rangle=\det\{\mathbf{P}^\dag(i)\mathbf{P}(j)\}.
\end{equation}
The matrices $\mathbf{P}(i)$ are obtained by taking the first $N$ rows from the matrix of the eigenvectors of $H_F$, then changing the sign of the first $i-1$ rows and adding a column with vanishing elements, except for the $i$-th one, which is equal to $1$. This provides an efficient way of computing $\rho_{ij}$ since it involves the calculation of determinants of $(N+1)\times(N+1)$ matrices instead of the more standard way that entails the evaluation of the Töplitz determinant of $(L+1)\times(L+1)$ matrices\cite{paredes04_tonks_gas,demartino05_disordered_bosons}.
One can take a step further and explicitly compute the product of matrices $\mathbf{P}^\dag(i)\mathbf{P}(j)$ for later reducing its determinant by using the property for block matrices
\begin{equation}
\det\begin{pmatrix}\mathbf{W} & \mathbf{X}\\
\mathbf{Y} & \mathbf{Z}
\end{pmatrix}=\det \mathbf{W}\,\det(\mathbf{Z}-\mathbf{Y}\mathbf{W}^{-1}\mathbf{X}),
\end{equation}
with $\mathbf{Z}$ the $1\times1$ matrix formed by the element $N+1,N+1$ of the product $\mathbf{P}^\dag(i)\mathbf{P}(j)$. The one-particle Green's function can thus be expressed as
\begin{equation}\label{eq:Gij}
G_{ij}=\sum_{m,n=1}^N\psi_m(i)A_{mn}(i,j)\psi_n^\ast(j),
\end{equation}
where $\psi_m(i)$ are the wavefunctions of $\mathbf{H}$. The $N\times N$ matrix $\mathbf{A}$ is
\begin{equation}\label{eq:Amat}
\mathbf{A}(i,j)=\mathbf{Q}^{-1}\det\mathbf{Q},
\end{equation}
where the entries of matrix $\mathbf{Q}$ are
\begin{equation}\label{eq:Qmat}
Q_{mn}(i,j)=\delta_{mn}-2\sum_{k=i}^{j-1}\psi^\ast_m(k)\psi_n(k).
\end{equation}
(for $j\geqslant i$). The advantage of Eq. (\ref{eq:Gij}) is that the numerical calculation of $G_{ij}$ for every pair of points can be optimized, and even parallelized, since $\mathbf{Q}(i,j+1)$ can be easily updated from $\mathbf{Q}(i,j)$. Notice that Eqs. (\ref{eq:Gij})-(\ref{eq:Qmat}) are the discrete version of the ones obtained in Ref. \onlinecite{pezer07_expression_OPDM} for a one-dimensional system of impenetrable bosons in the continuum.

At finite temperature $T$, Eq. (\ref{eq:opdm}) is obtained from a thermal average over a the grand-canonical ensemble described by a density operator $\rho=e^{-(H-\mu \hat{N})/\kB  T}/Z$, where $Z=\tr e^{-(H-\mu \hat{N})/\kB  T}$. As usual, the chemical potential $\mu$ is determined implicitly by the condition that fixes the mean particle number $N=\sum_i \rho_{ii}$. For the calculation of the nondiagonal elements we first employ the method proposed in Ref. \onlinecite{rigol05_hcbosons_finiteT} that makes use of the properties of Slater determinants and leads to the following result for the  one-particle density matrix:
\begin{multline}\label{eq:RigolResult}
\rho_{ij}={\frac{1}{Z}}\{\det[\textbf{I}+(\textbf{I}+\textbf{A})\textbf{O}(j)\textbf{U}\mathbf{R}\textbf{U}^{\dag}\textbf{O}(i)] \\ -\det[\textbf{I}+\textbf{O}(j)\textbf{U}\mathbf{R}\textbf{U}^{\dag}\textbf{O}(i)]\},
\end{multline}
where $\mathbf{U}$ is the unitary matrix containing the eigenvector of $\mathbf{H}$ in its columns, $\mathbf{O}(i)$ is diagonal with the first $i-1$ elements of the diagonal equal to $-1$ and the others equal to $1$, and the matrix $\mathbf{A}$ has only one non-zero element: $A_{ij}=1$. In addition, we defined $\mathbf{R}=e^{-(\textbf{E}-\mu\textbf{I})/\kB T}$ with $\mathbf{I}$ the identity matrix and $\mathbf{E}=\mathbf{U}^{\dag}\mathbf{H}\mathbf{U}$ the diagonal matrix with the eigenvalues of $\mathbf{H}$.

However, the evaluation of these determinants can be unstable at low temperatures since the matrix $\mathbf{R}$ contains very small and very large elements in its diagonal. In order to obtain more stable numerical results, it is convenient to isolate $\mathbf{R}$ by rearranging the factors inside the first term of Eq. (\ref{eq:RigolResult}) as
\begin{multline}\label{eq:large_small_ev}
\det\left[\mathbf{I}+\left(\mathbf{I}+\mathbf{A}\right)\mathbf{O}(j)\mathbf{U}\mathbf{R}\mathbf{U}^{\dagger}\mathbf{O}(i)\right]
=\det\mathbf{O}(j)\mathbf{O}(i)\\
\times\det\left[\mathbf{U}^{\dagger}\mathbf{O}(j)\mathbf{O}(i)\mathbf{U}+\mathbf{R}+\mathbf{U}^{\dagger}\mathbf{A}\mathbf{U}\right],
\end{multline}
and the same for the second term of Eq. (\ref{eq:RigolResult}). Moreover, we notice that the product $\mathbf{U}^\dagger\mathbf{A}\mathbf{U}$
can be written as an outer product,
\begin{equation}
[\mathbf{U}^{\dagger}\mathbf{A}\mathbf{U}]_{mn}=\psi^\ast_m(i)\psi_n(j).
\end{equation}
Therefore, we can apply to the right-hand side of Eq. (\ref{eq:large_small_ev}) the matrix determinant lemma
\begin{equation}
\det\left(\mathbf{M}+\mathbf{u}\otimes\mathbf{v}^{T}\right)=\det\mathbf{M}\left(1+\mathbf{v}^{T}\mathbf{M}^{-1}\mathbf{u}\right)
\end{equation}
for any invertible square matrix $\mathbf{M}$, and we finally obtain for the one-particle density matrix the result
\begin{equation}\label{eq:rho}
\rho_{ij}=\sum_{m,n=1}^L\psi^\ast_m(i)B_{mn}(i,j)\psi_n(j),
\end{equation}
where now the $L\times L$ matrix $\mathbf{B}$ is
\begin{equation}\label{eq:Bmat}
\mathbf{B}(i,j)=(-1)^{j-i}\frac{\det(\mathbf{Q}+\mathbf{R})}{\det(\mathbf{I}+\mathbf{R})}(\mathbf{Q}^T+\mathbf{R})^{-1},
\end{equation}
and $\mathbf{Q}$ is the extension of the matrix defined in Eq. (\ref{eq:Qmat}) to size $L\times L$. In order to evaluate the one-particle density matrix at finite temperature, one needs to calculate the $L\times L$ matrix $\mathbf{Q}+\mathbf{R}$, which can be updated from site to site, and its inverse and determinant for each pair of sites, for which there are fast and accurate numerical procedures. Equations. (\ref{eq:rho}) and (\ref{eq:Bmat}) are the finite-temperature analogs of Eqs. (\ref{eq:Gij})-(\ref{eq:Qmat}).

\section{Finite temperature crossover}

\begin{figure}
\begin{center}
\includegraphics [width=\figwidth]{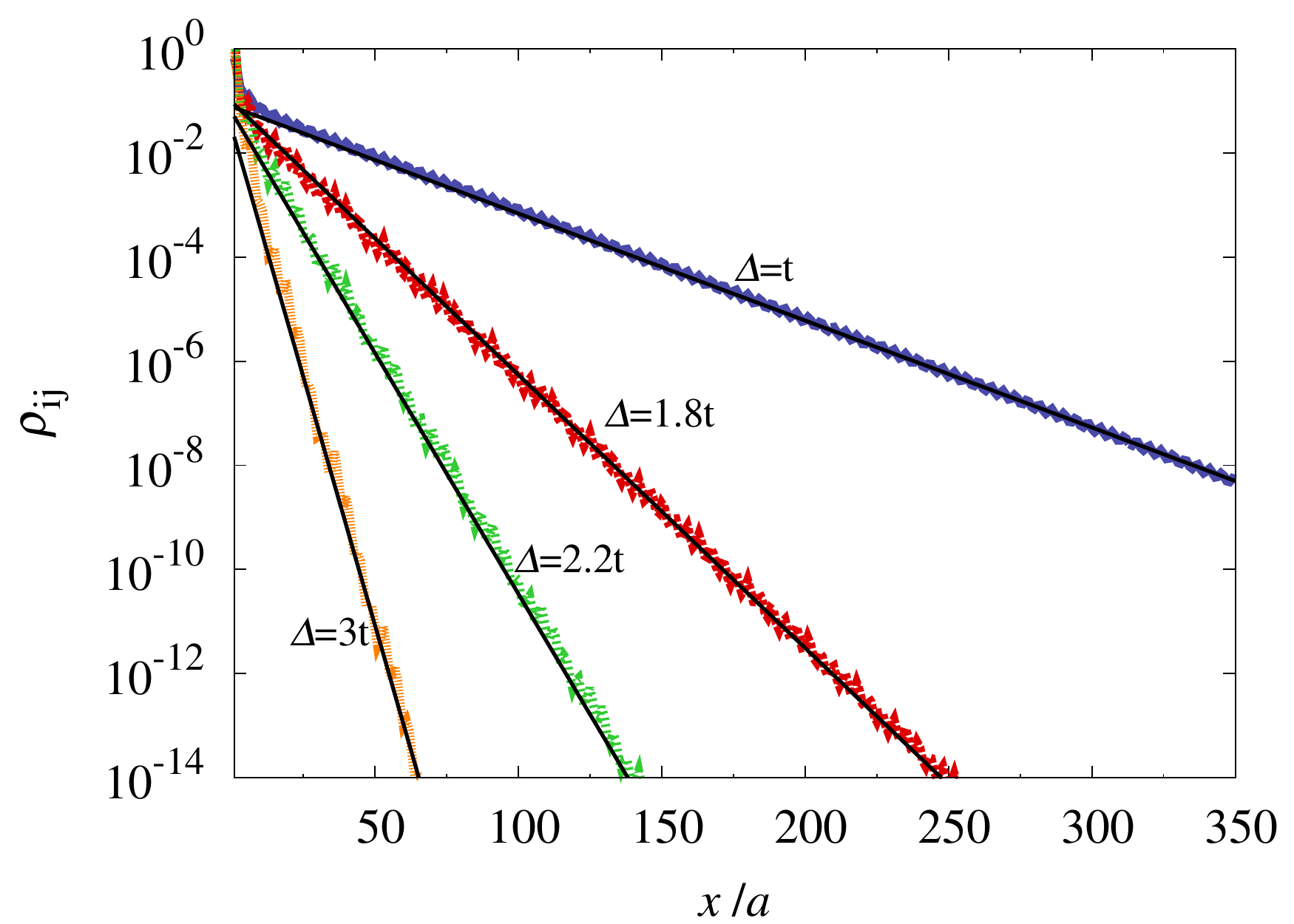}
\caption{\label{opdm-decay}Log-log plots of the decay of the one-particle density matrix with distance (in units of the lattice constant $a$) for $\kB T=0.04t$ and different values of $\Delta$ in a half-filled systems with 1000 lattice sites. Thin solid lines are exponential fits.}
\end{center}
\end{figure}

At zero temperature, the phase diagram of the Hamiltonian described by Eq. (\ref{eq:hamhcb}) has been extensively studied \cite{roscilde08_one_dimensional_superlattices,%
roux08_quasiperiodic_phase_diagram,%
deng08_phase_diagram_bichromatic,%
deng09_superfluid_quasiperiodic,%
cai10_hc-bosons_quasi-periodic}, and it was found that three phases exist, namely, superfluid, Bose-glass, and incommensurate insulator, depending on the value of the chemical potential and $\Delta$. One can relate this zero-temperature phase diagram to the single-particle band structure of the fermionic Hamiltonian of Eq. (\ref{eq:fercuasids}). The energy spectrum  of $H_F$  is organized in a series of bands~\cite{aubry80_andre_model,sokoloff85_review_harper_equation}. Its wave functions are all localized for $\Delta>\Delta_c$, with $\Delta_c=2t$, and all extended for $\Delta<\Delta_c$ giving rise to the Bose-glass and the superfluid phases, respectively, when the chemical potential lies inside one of the bands. For values of the chemical potential inside a gap, the system is an incommensurate insulator for any value of $\Delta$. By fixing the chemical potential in the middle of a band, one can study the transition from the Bose glass to the superfluid by varying $\Delta$. We explicitly analyze this case by setting $\mu=0$ in all the calculations which fixes the mean number of particles to half filling.

At zero temperature the one-particle density matrix exhibits a power law decay $\rho_{ij}\propto |x_i-x_j|^{-\eta}$ for $\Delta<\Delta_c$, characteristic of a quasi-long-range ordered superfluid phase. In the absence of the quasiperiodic potential, the exponent is $\eta=1/2$~\cite{mccoy68_power_law_XY}. As the incommensurate secondary potential is turned on $\eta$ becomes smaller, wwhich is accompanied by a quick decrease of the superfluid fraction~\cite{cai10_hc-bosons_quasi-periodic}. When $\Delta$ exceeds the critical point $\Delta_c$, the one-particle density matrix decay turns into an exponential law $\rho_{ij}\sim e^{-|x_i-x_j|/\xi}$, where $\xi$ is the correlation length, signaling the transition to the disordered Bose-glass phase~\cite{fisher89_bose_glass}. In one-dimensional (1D) systems any finite temperature destroys quasi-long-range order, inducing an exponential decay of the correlations characterized by a temperature- (and $\Delta$ in the problem under study) dependent correlation length. For hard-core bosons in the absence of the quasiperiodic potential this has been verified earlier~\cite{rigol05_hcbosons_finiteT}, where the power-law behavior of the correlation length with increasing temperature $\xi\sim 1/T$ was found for low temperatures $\kB  T < t$. Here we extend these calculations to finite $\Delta$, investigate the finite-temperature crossovers associated with the quantum critical point at $\Delta=\Delta_c$~\cite{cai11_quantum_critical}, and characterize the properties of the different regions in the $T-\Delta$ phase diagram.

\begin{figure}[t]
\begin{center}
\includegraphics [width=\figwidth]{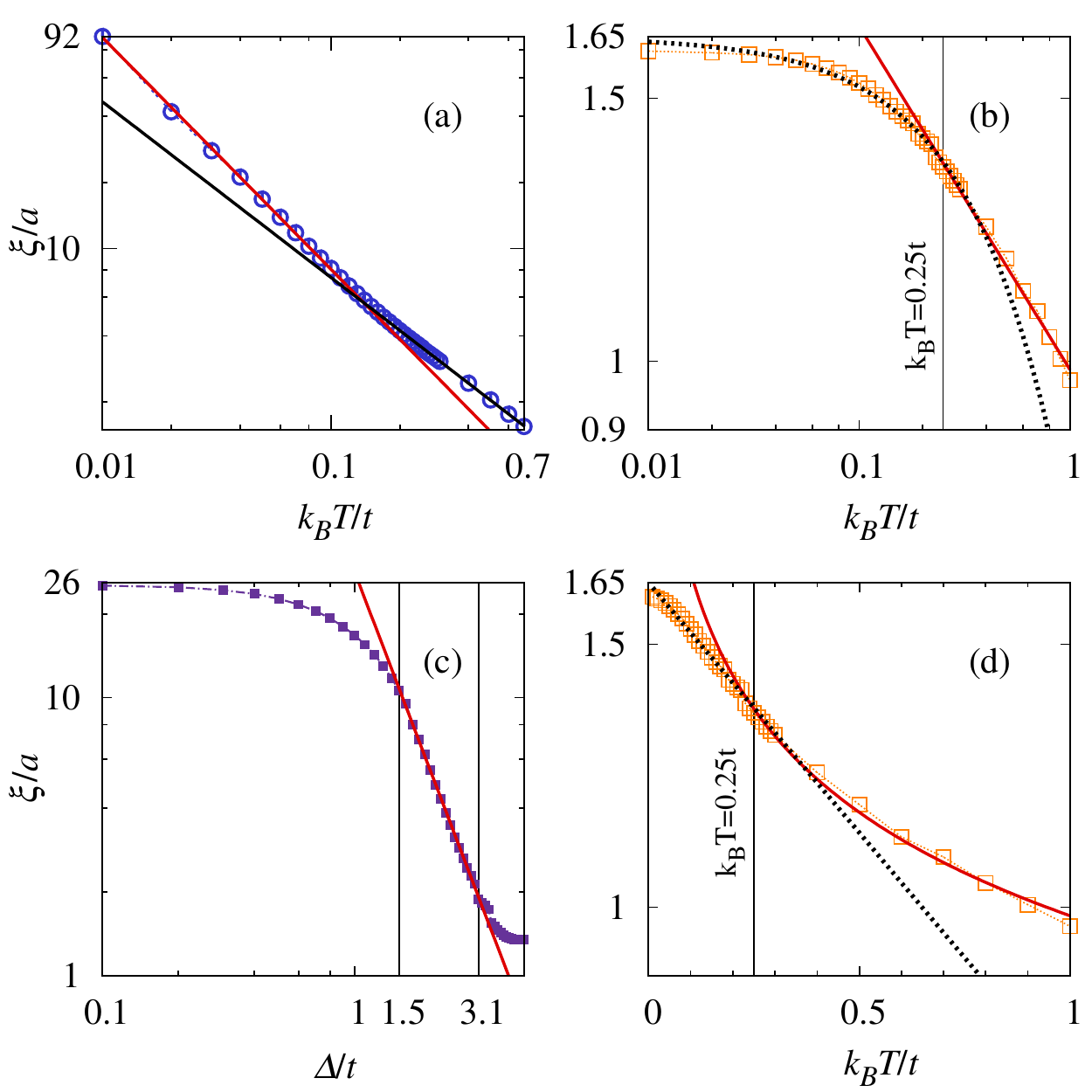}
\caption{\label{corrlen}Correlation length as a function of temperature and $\Delta$. (a) $\xi(T)$ for $\Delta=t$ (in logarithmic scale). Solid thin lines show power law fits. (b) $\xi(T)$ for $\Delta=3.5t$ in log-log scale. A power-law behavior is observed for $\kB T>0.25t$. (c) $\xi(\Delta)$ for $\kB T=0.05t$. (d) Same as (b) in log-linear scale.}
\end{center}
\end{figure}

\begin{figure}
\begin{center}
\includegraphics [width=\figwidth]{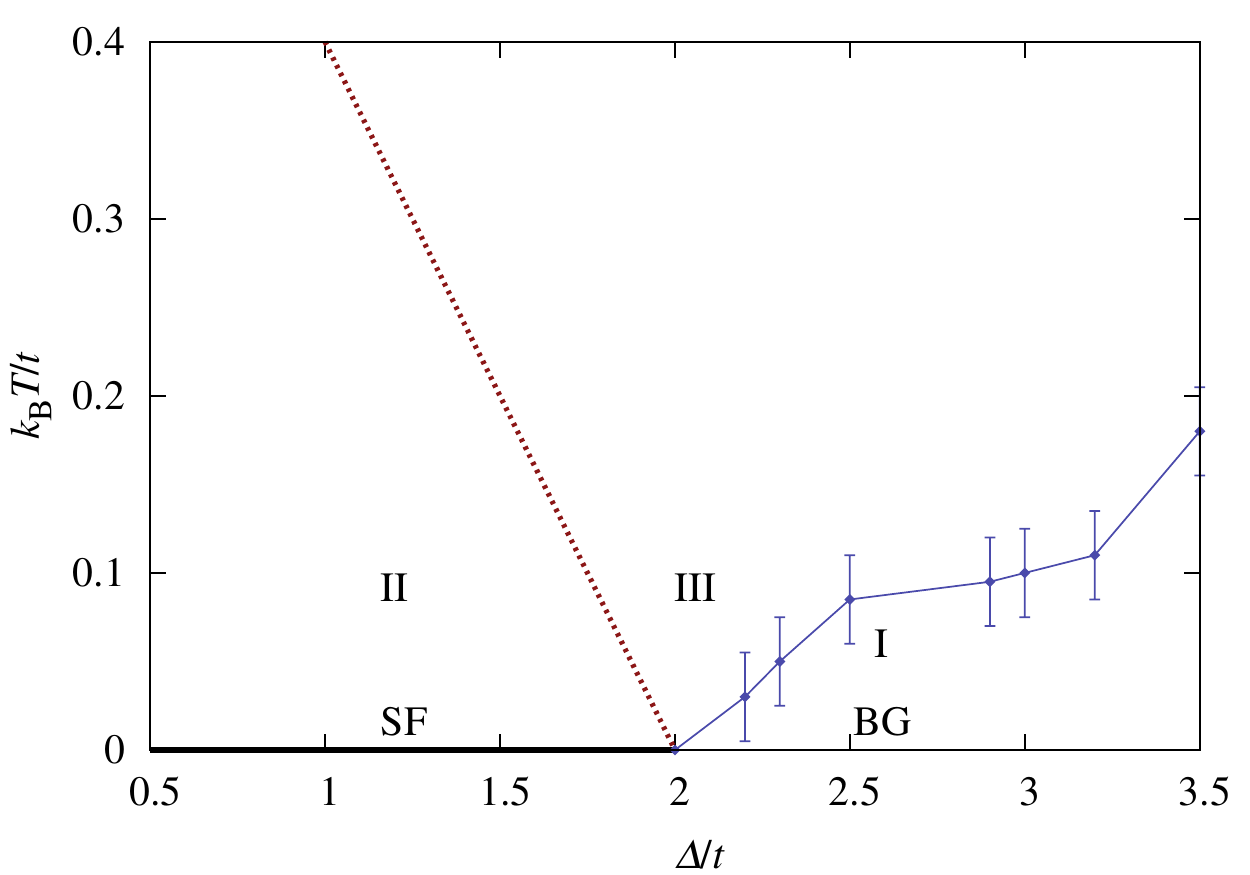}
\caption{\label{phase}Finite $T$ phase diagram of 1D hard core bosons in a quasiperiodic optical lattice at half-filling in the $\Delta-T$ space in the vicinity of the quantum critical point at $\Delta=\Delta_c=2t$. Quasi-long-range order is present at $T=0$ for $\Delta<\Delta_c$ as a superfluid (SF). The ground state for $\Delta>\Delta_c$ is a Bose glass (BG). Regions I, II, and III are characterized by the temperature dependence of the correlation length. We sketch the phase boundary between regions II and III.}
\end{center}
\end{figure}

Figure~\ref{opdm-decay} shows the decay of the one-particle density matrix for a half-filled system of 1000 lattice sites at a finite temperature $\kB  T = 0.04 t$ and different values of $\Delta$. The decay is exponential, characterized by a correlation length $\xi$ whose temperature and $\Delta$ dependence is shown in Fig.~\ref{corrlen}. The study of the behavior of $\xi$ for different regimes allows us to establish the crossover phase diagram shown in Fig.~\ref{phase}. For values of $\Delta\leqslant \Delta_c$, we find that the correlation length diverges at low temperatures as a power law $\xi\sim T^{-\mu_-}$ [Fig.~\ref{corrlen}(a)], with an exponent $\mu_-$ that depends on $\Delta$. This is marked as region II in Fig.~\ref{phase} and can be identified with a thermally disordered phase in which quasi-long-range superfluid order is destroyed by thermal fluctuations. We also observe a crossover to a power law behavior with a different exponent $\mu_+$ for temperatures above a crossover temperature $T_-(\Delta)$, (region III in Fig.~\ref{phase}). The obtained data allows us to make a sketch of the phase border between regions II and III. On the localized side of the spectrum the correlation length behaves as a power law with increasing temperature for $T$ larger than the right branch of the crossover temperature $T_+(\Delta)$ [Fig~\ref{corrlen}(b)], but when lowering the temperature, it saturates to the zero-temperature finite value following the exponential behavior $\xi\sim e^{-a T}$ for $T<T_+(\Delta)$ [Fig~\ref{corrlen}(d)]. This is represented in Fig.~\ref{phase} as region I, and can be identified as a quantum disordered phase, since its finite-temperature properties are dictated by the localized Bose-glass ground state. In particular, as we see in the next section, natural orbitals are still localized in this regime. Region III, in between I and II, is essentially the quantum critical region in which both types of fluctuations, quantum and thermal are important. A more precise determination of the phase borders, in particular between phase II and III, will be done elsewhere~\cite{nessi_unpublished}.

\begin{figure}
\begin{center}
\includegraphics [width=\figwidth]{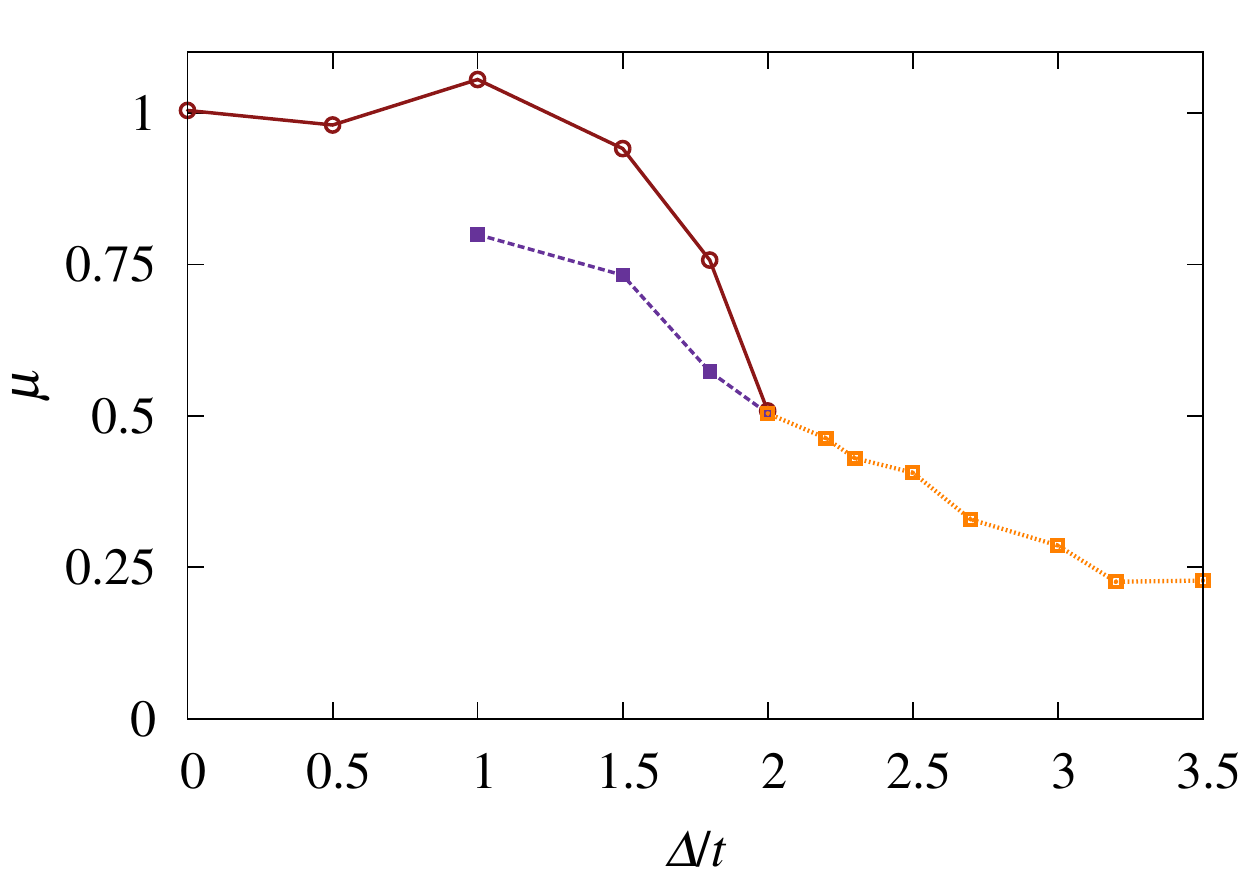}
\caption{Estimated exponents of the power-law behavior of the correlation length, $\xi\sim T^{-\mu}$, as a function of $\Delta$. For $\Delta<\Delta_c$ we show the low-temperature exponent $\mu_-$ in brown (solid) and the higher-temperature exponent  $\mu_+$ in blue (dashed). In orange (dotted) we represent the high-temperature exponent for $\Delta>\Delta_c$. Notice that at $\Delta=\Delta_c$ both low- and high-temperature exponents tend to converge to a common value.\label{mu}}
\end{center}
\end{figure}

Figure~\ref{mu} shows the nonuniversal behavior of the exponents $\mu_\pm$ as a function of $\Delta$ for the whole range of values of $\Delta$ and the two temperature regimes when $\Delta<\Delta_c$. For $\Delta < 1.5t$ the exponent $\mu_-$ is almost constant retaining the zero $\Delta$ value $\mu_-=1$ whereas for $\Delta>1.5t$ it decays abruptly to its limit value at $\Delta=\Delta_c$ (for $\Delta>\Delta_c$ the low-temperature behavior changes to an exponential form). On the other hand, the high-temperature exponent $\mu_+$ shows a smoother decreasing behavior with $\Delta$, that continues to the $\Delta>\Delta_c$ region. At $\Delta=\Delta_c$, we still have two regimes with the same exponent $\mu_+\simeq \mu_-\simeq 1/2$ that differ in the prefactor. Even though the low-temperature region is very small, we should expect a single high-temperature regime at the critical point. We attribute the survival of the low-temperature one to a finite size effect.

Figure~\ref{corrlen}(c) shows the dependence of $\xi$ on $\Delta$ for $\kB T=0.05t$. For low $\Delta$ the correlation length converges to the zero disorder value. As $\Delta$ is raised the correlation length smoothly switches to a power law at a certain value of the secondary lattice amplitude that moves to smaller values for increasing temperature. For even higher amplitudes of the incommensurate potential we find a new crossover, and the behavior departs from a power law.

The behavior of the one-particle density matrix for $\Delta\lesssim 1$ and fillings such that we are far from the band edges can be understood in terms of the Luttinger liquid description of the 1D interacting bosonic fluid, using a low-energy hydrodynamic approach~\cite{giamarchi04_book_1d,haldane81_effective_harmonic_fluid_approach,cazalilla04_bosonization_1D_cold_atoms,roux08_quasiperiodic_phase_diagram,deng08_phase_diagram_bichromatic}. In particular, the order parameter correlation function in the thermodynamic limit (and in the continuum) is given by~\cite{cazalilla04_bosonization_1D_cold_atoms}
\begin{equation}
\rho(x,x')\sim\left(\frac{\pi/L_T}{\sinh(\pi |x-x'|/L_T)}\right)^{1/2K},
\end{equation}
where $L_T\sim 1/T$ is the thermal correlation length and $K$ is the Luttinger liquid parameter, which takes the value $K=1$ in the absence of the secondary lattice, but is renormalized at finite $\Delta$. From this result one observes that correlations cross over from algebraic decay for $|x-x'|<<L_T$ with exponent $1/2K$ to an exponential form for $|x-x'|>>L_T$. The latter case results in a behavior $\xi\sim 1/T$ in agreement with Ref. \onlinecite{rigol05_hcbosons_finiteT} and the results of Fig. \ref{mu}. The fact that the exponent $\mu$ departs from the value $\mu=1$ for $\Delta\gtrsim t$ signals the breakdown of the Luttinger liquid description as the system approaches the quantum critical region.

\begin{figure}[t]
\begin{center}
\includegraphics [width=\figwidth]{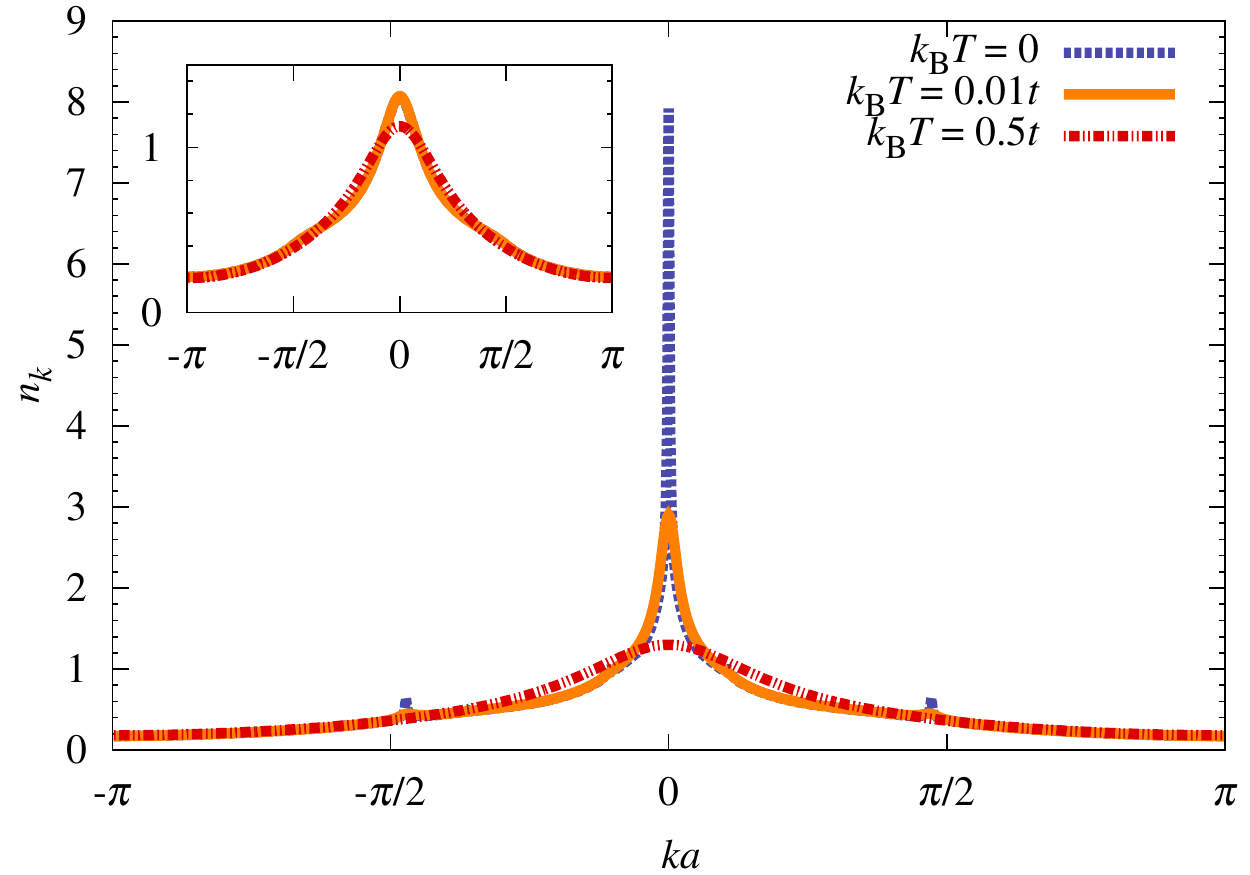}
\caption{\label{nk}Momentum distribution for $\Delta=1.9t$ (main plot) and $\Delta=2.5t$ (inset) at different temperatures for a half-filled system of 800 lattice sites. At $\Delta=2.5t$ curves for $\kB T=0$ and $\kB T = 0.01t$ are identical.}
\end{center}
\end{figure}

Let us finally analyze the momentum distribution function,
\begin{equation}
n_k=\frac{1}{L}\sum_{ij}e^{-ik(x_i-x_j)}\rho_{ij},
\end{equation}
where $k$ denotes the momentum. In Fig. \ref{nk} we show the behavior of the momentum distribution for two representative values of $\Delta$ in a system of $800$ lattice sites. A finite temperature reinforces the effect of the pseudodisorder on the correlations. At $\Delta=t$, temperature washes out the two fundamental features of the momentum distribution: the central peak (associated with the quasi-long-range correlations at $T=0$) and the secondary peaks (related to the secondary incommensurate lattice). In the absence of the secondary lattice, at $\kB  T=0.01t$ the height of the central momentum peak was found to decrease to 2/3 of the one at zero temperature~\cite{rigol05_hcbosons_finiteT}. A finite value of $\Delta=1.9t$ at the same temperature reduces even further its height to around 1/3 of the central one, though the secondary peaks are still visible. At $\kB  T=0.5t$ both features essentially disappeared.

\section{Delocalization of the natural orbitals}

Since the single-particle wavefunctions of the Hamiltonian (\ref{eq:fercuasids}) are all either localized for $\Delta>\Delta_c$ or delocalized for $\Delta<\Delta_c$, that is, there are no mobility edges, one expects activation effects to be suppressed and temperature to play no role in the physics of delocalization. However, the Jordan-Wigner transformation leads to a highly complicated nonlocal form of the bosonic operators, which keeps track of the strong interactions, and might account for nontrivial effects in the vicinity of a quantum critical point, hidden in the apparent simplicity of the fermionic Hamiltonian. We analyze here these effects by studying the localization properties of the natural orbitals defined as the eigenvectors of the one-particle density matrix:
\begin{figure}[t]
\begin{center}
\includegraphics [width=\figwidth]{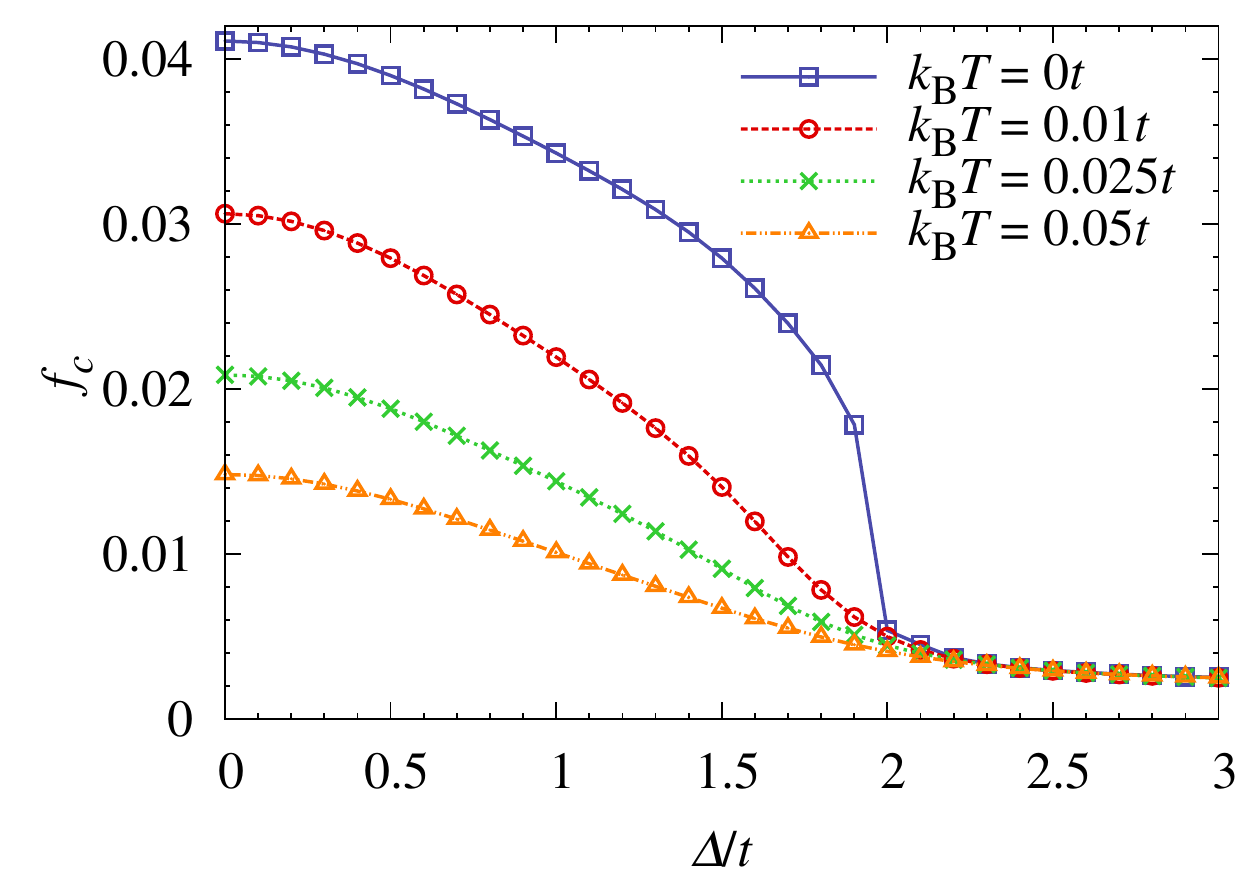}
\caption{\label{c_frac}Condensate fraction as a function of $\Delta$ for different temperatures. At $\Delta=\Delta_c$ it reaches a common residual value signaling the transition to the Bose glass.}
\end{center}
\end{figure}
\begin{figure}[b]
\begin{center}
\includegraphics [width=\figwidth]{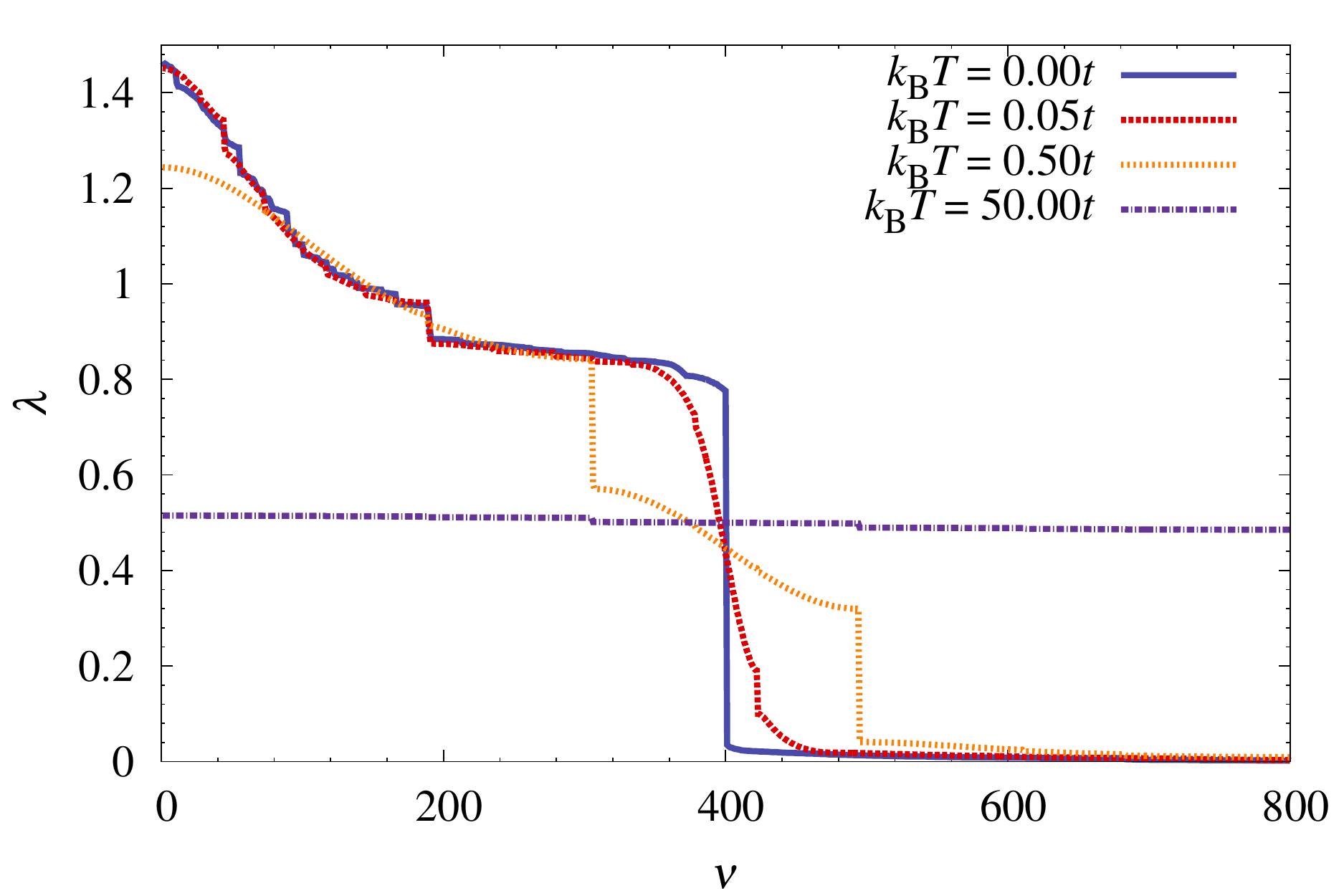}
\caption{\label{occ}Occupations of the natural orbitals at $\Delta=2.5t$ and different temperatures for a half-filled system of 800 lattice sites. The abrupt jump observed at $\nu=400$ at zero temperature is rounded as temperature rises. For very high temperatures all the orbitals are equally occupied.}
\end{center}
\end{figure}
\begin{figure*}[t]
\begin{center}
\includegraphics [width=0.99\textwidth]{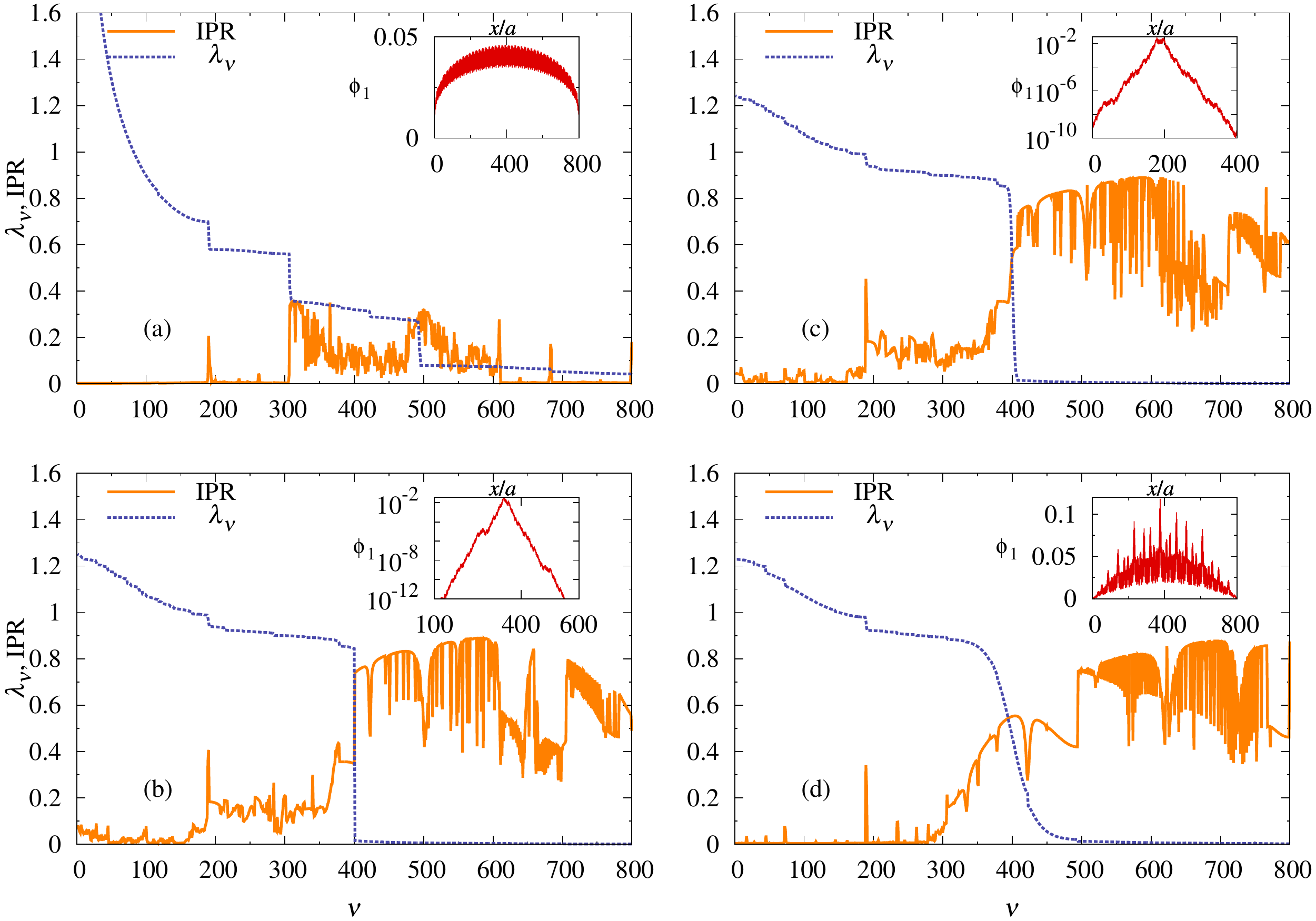}
\caption{\label{ipr} Occupation and IPR of the orbitals at (a) $\kB T=0$ and $\Delta=t$, (b) $\kB T=0$ and $\Delta=3t$, (c) $\kB T=0.01t$ and $\Delta = 3t$, (d) $\kB T=0.1t$ and $\Delta = 3t$ for a half-filled system of 800 lattice sites. In the insets we show the profile of the first orbital. Note the log scale in (b) and (c).}
\end{center}
\end{figure*}
\begin{equation} \label{eq:natorb}
\sum_{j=1}^{N}\rho_{ij}\phi_{j}^{\nu}=\lambda_{\nu}\phi_{i}^{\nu}.
\end{equation}
Here the corresponding eigenvalues \{$\lambda^{\nu}$\} represent the occupations. The natural orbitals can be considered as effective one-particle states~\cite{penrose56_BEC_he4} in systems with interactions. For noninteracting systems the natural orbitals reduce to the exact independent-particle eigenfunctions of the Hamiltonian, and at finite temperature their occupancies are either the Bose or Fermi factor depending on the character of the statistics. However, when interactions are present as in Eq. (\ref{eq:hamhcb}), and unlike the eigenstates of the Hamiltonian (\ref{eq:fercuasids}), the natural orbitals depend on temperature~\cite{rigol05_hcbosons_finiteT}. In dilute high-dimensional gases exhibiting Bose-Einstein condensation (BEC), the lowest natural orbital is usually regarded as the condensate wave function, and the BEC order parameter~\cite{leggett01_review_BEC}. In systems with disorder, the phenomenon of Anderson localization in dilute gases is usually studied by considering the localization of the condensate wave function~\cite{larcher09_localization_GP_1D}. Here we extend this analysis to higher natural orbitals since they turn out to be occupied when the system approaches the Bose-glass phase.

Let us investigate first the occupations of the natural orbitals. The superfluid phase at zero temperature is associated with the formation of a quasi-condensate, that is, a large
occupation of the lowest natural orbitals can occur. For instance, for hard-core bosons in the absence of the secondary lattice, $\lambda_1$ scales as $\sim\sqrt{N}$~\cite{rigol04_lattice_hcbosons}. In Fig.~\ref{c_frac} we observe the occupation of the first orbital for different temperatures along with the $T=0$ results as a function of the incommensurate potential amplitude. For all temperatures the condensate occupation decreases with increasing $\Delta$, and reaches a common residual value at $\Delta=\Delta_c$ the value at which the single-particle eigenfunctions localize and the system enters the Bose-glass. Another signature of the Bose-glass phase at zero temperature is the opening of a discontinuity in the occupation of the orbitals for $\nu=N$, resembling the Fermi-Dirac distribution, as can be observed in Fig.~\ref{occ}, in which we show the distribution of occupations of the orbitals versus the orbital number for fixed $\Delta$ at different temperatures. However, the analogy with the Fermi-Dirac distribution function is not complete, since the occupations of natural orbitals below a certain threshold (of around $\nu=N/2$) become larger than 1 and slowly increase when descending in the orbital number. The discontinuity at $\nu=N$ is erased by thermal effects even at very low temperatures. However, at higher temperatures ($\kB T\simeq 0.5t$ for $\Delta=2.5t$) two discontinuities develop symmetrically placed around $\nu=N$. At very high temperatures we approach the infinite-temperature regime, in which all occupations are equal to $1/2$.

At zero temperature, the superfluid-to-Bose-glass transition is marked by the localization of the lowest orbitals at $\Delta=\Delta_c$. The insets of Figs.~\ref{ipr}(a) and~\ref{ipr}(b) show the effect of the incommensurate potential on the lowest orbital below and above the transition. For $\Delta<\Delta_c$ the states are extended, whereas for $\Delta>\Delta_c$ they exponentially localize, which mimics the behavior of the single-particle states of the fermionic Hamiltonian of Eq. (\ref{eq:fercuasids}). Nevertheless, as one approaches the point $\Delta=\Delta_c$ the range of significantly occupied high-order orbitals increases, implying that the dominance of the lowest ones decreases. Therefore, the characterization of the system by a single macroscopic wave function $\phi_1$ can be put into question with increasing $\Delta$, and one is led to analyze the whole set of significantly occupied orbitals. Given a single-particle state $|\phi\rangle$ we can quantify its degree of localization by the inverse participation ratio (IPR) in real space, defined as
\begin{equation}\label{eq:ipr}
 \text{IPR}(\phi)\equiv \sum^L_{j=1} |\langle\phi|j\rangle|^4
\end{equation}
where $|j\rangle$ is the state at site $j$ in the real-space basis. The IPR is a measure of the inverse number of sites occupied by the state as can be understood by regarding two limiting cases. If the state is localized on a single site $i$, we immediately find from $\langle\phi|j\rangle=\delta_{ij}$ that $\text{IPR}=1$. If, on the other hand, the state is distributed equally over $L$ sites, we have $\langle\phi|j\rangle=L^{-1/2}$ and thus $\text{IPR} = 1/L$.

Figures~\ref{ipr}(a) and ~\ref{ipr}(b) show the IPR for the whole set of natural orbitals together with their occupation for two values of $\Delta$, below and above the transition at zero temperature. We observe that for $\Delta<\Delta_c$, where all the single-particle fermionic states are extended, there is a group of localized natural orbitals between $\nu=300$ and $\nu=600$ for a half filled system of $800$ lattice sites. For $\Delta>\Delta_c$, in which case all single-particle fermionic wave functions are localized, some natural orbitals are extended between $\nu=50$ and $\nu=150$.

\begin{figure}
\begin{center}
\includegraphics [width=\figwidth]{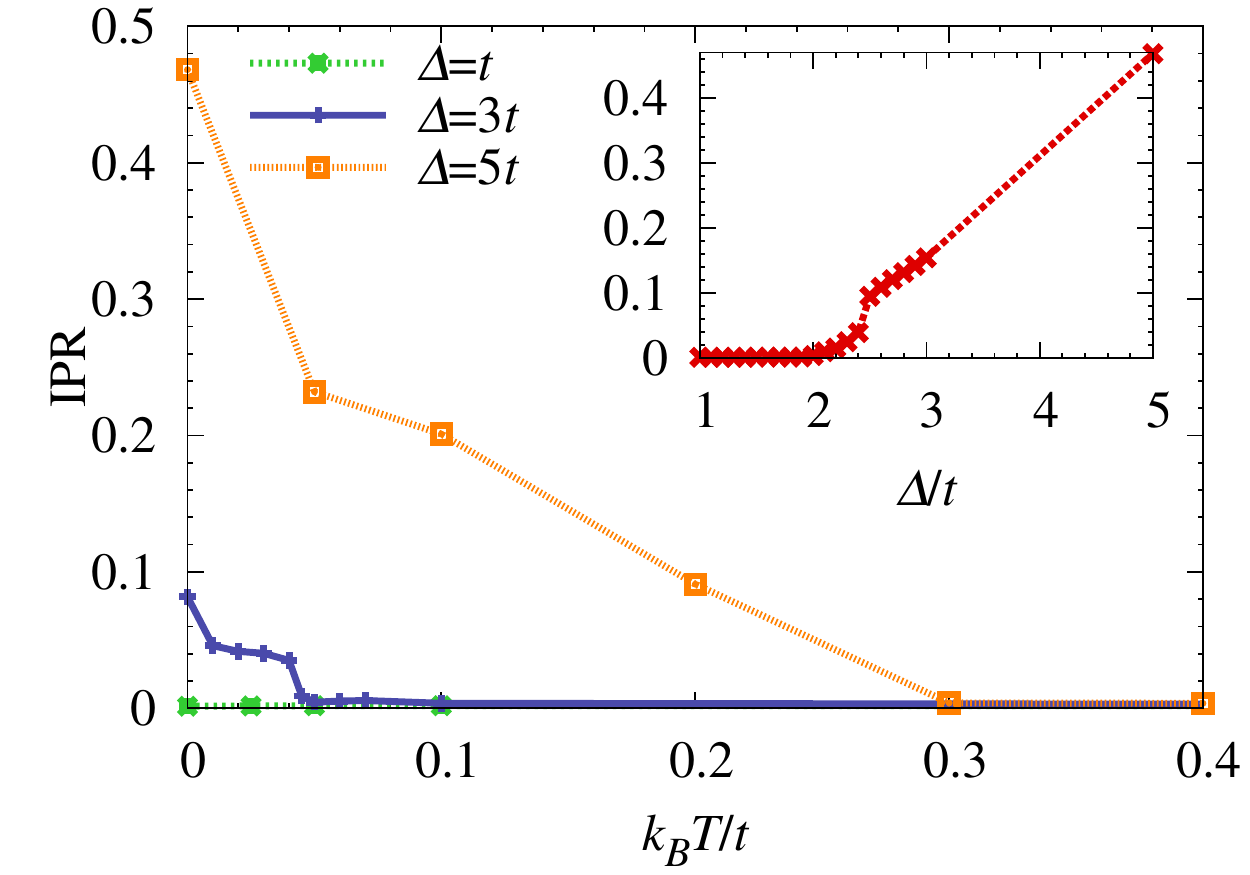}
\caption{\label{iprtd}Inverse participation ratio of the first orbital as a function of temperature for different values of the secondary lattice amplitude. (Inset) $\Delta$ dependence of the IPR for the first orbital at $T=0$.}
\end{center}
\end{figure}

The effect of a finite temperature on the natural orbitals is shown in Figs.~\ref{ipr}(c) and ~\ref{ipr}(d) where the profile of the first natural orbitals for $\Delta=3t$ at two different temperatures is plotted in the insets. As the temperature is increased from zero to a small but finite value, the lowest natural orbitals are still exponentially localized. This is in agreement with the fact that in this finite-temperature region, the physics is dominated by the quantum fluctuations of the Bose-glass ground state. When the temperature exceeds some value $T_d=T_d(\Delta)$, of about $T_d\approx0.05t$ for $\Delta = 3t$ the lowest natural orbitals are not localized anymore; instead they spread over all the lattice sites, even for $\Delta>\Delta_c$. This suggests a sort of competition between the effects of the incommensurate potential that acts like a random potential on sites because of the irrational value of $\sigma$ and tends to localize the natural orbitals and thermal effects traceable solely trough correlations of nonlocal operators for which the system is fully interacting, that aims to delocalize them. Nevertheless, as in the zero-temperature case, not all the orbitals exhibit the same behavior, not even all the occupied ones. Figure~\ref{ipr} also shows the occupations and the IPR of all the natural orbitals, and we observe the same situation as at zero temperature: The delocalization occurs for orbitals such that $\nu\lesssim 300$, whereas the group of orbitals around and above the step in the occupations at $\nu=N$ stay localized. Finally, we show in Fig.~\ref{iprtd} the IPR of the first orbital as a function of temperature for different values of $\Delta$. For the two values of $\Delta>\Delta_c$ shown, the IPR reduces to a residual value when the temperature exceeds $T_d(\Delta)$. Remarkably, the temperature $T_d$ of delocalization of the lowest orbitals is very close to the crossover temperature $T_+$ at which the correlation length changes its behavior as a function of temperature.

\section{Summary}

In conclusion, we have studied finite temperature properties of hard-core bosons in a 1D incommensurate lattice potential. Based on the Jordan-Wigner transformation and by using properties of Slater determinants, we computed the one-particle density matrix and related quantities such as the momentum distribution and natural orbitals for a half-filled system. We obtained a finite-temperature crossover phase diagram in which different regions are classified according to the behavior of the correlation length with temperature. For values of the incommensurate potential amplitude below the critical value $\Delta_c$ we find that the correlation length diverges at low temperatures as a power law with a $\Delta$-dependent exponent. As the system enters into the quantum critical region, the correlation-length crosses over to a power-law form with a different exponent. In contrast, for $\Delta>\Delta_c$ there is a crossover from a low-temperature exponential behavior of the correlation length, characteristic of the quantum fluctuations in the Bose-glass ground state, to the power-law form when the systems enters into the critical region on the localized side of the spectrum. We also found evidence that this crossover is characterized by an important change in the localization properties of the natural orbitals, in particular a significative fraction of the natural orbitals delocalize when the temperature exceeds the crossover temperature scale.

\acknowledgments

This work was supported by grants from CONICET (PIP 0662), ANPCyT (PICT 2010-1907) and UNLP (PID X497), Argentina.

\end{document}